\documentclass[12pt]{article}
\usepackage{amssymb}
\usepackage{rotating}

\begin{document}



\def\a{\alpha}
\def\b{\beta}
\def\d{\delta}
\def\e{\epsilon}
\def\g{\gamma}
\def\h{\mathfrak{h}}
\def\k{\kappa}
\def\l{\lambda}
\def\o{\omega}
\def\p{\wp}
\def\r{\rho}
\def\t{\tau}
\def\s{\sigma}
\def\z{\zeta}
\def\x{\xi}
\def\V={{{\bf\rm{V}}}}
 \def\A{{\cal{A}}}
 \def\B{{\cal{B}}}
 \def\C{{\cal{C}}}
 \def\D{{\cal{D}}}
\def\G{\Gamma}
\def\K{{\cal{K}}}
\def\O{\Omega}
\def\R{\bar{R}}
\def\T{{\cal{T}}}
\def\L{\Lambda}
\def\f{E_{\tau,\eta}(sl_2)}
\def\E{E_{\tau,\eta}(sl_n)}
\def\Zb{\mathbb{Z}}
\def\Cb{\mathbb{C}}

\def\R{\overline{R}}

\def\beq{\begin{equation}}
\def\eeq{\end{equation}}
\def\bea{\begin{eqnarray}}
\def\eea{\end{eqnarray}}
\def\ba{\begin{array}}
\def\ea{\end{array}}
\def\no{\nonumber}
\def\le{\langle}
\def\re{\rangle}
\def\lt{\left}
\def\rt{\right}

\newtheorem{Theorem}{Theorem}
\newtheorem{Definition}{Definition}
\newtheorem{Proposition}{Proposition}
\newtheorem{Lemma}{Lemma}
\newtheorem{Corollary}{Corollary}
\newcommand{\proof}[1]{{\bf Proof. }
        #1\begin{flushright}$\Box$\end{flushright}}

\baselineskip=20pt

\newfont{\elevenmib}{cmmib10 scaled\magstep1}
\newcommand{\preprint}{
   \begin{flushleft}
   \end{flushleft}\vspace{-1.3cm}
   \begin{flushright}\normalsize
   \end{flushright}}
\newcommand{\Title}[1]{{\baselineskip=26pt
   \begin{center} \Large \bf #1 \\ \ \\ \end{center}}}

\newcommand{\Author}{\begin{center}
   \large \bf
Guang-Liang Li${}^{a}$, Junpeng Cao${}^{b,c,d}$, Panpan Xue${}^{a}$,
Zhi-Rong Xin${}^{e}$,  Kun Hao${}^{e,f}$, Wen-Li
Yang${}^{e,f,g}\footnote{Corresponding author: wlyang@nwu.edu.cn}$
, Kangjie Shi${}^e$ and Yupeng
Wang${}^{b,d}\footnote{Corresponding author: yupeng@iphy.ac.cn}$
 \end{center}}

\newcommand{\Address}{\begin{center}

     ${}^a$Department of Applied Physics, Xian Jiaotong University, Xian 710049, China\\
     ${}^b$Beijing National Laboratory for Condensed Matter
           Physics, Institute of Physics, Chinese Academy of Sciences, Beijing
           100190, China\\
     ${}^c$Songshan Lake Materials Laboratory, Dongguan, Guangdong 523808, China \\
     ${}^d$School of Physical Sciences, University of Chinese Academy of
Sciences, Beijing, China\\
     ${}^e$Institute of Modern Physics, Northwest University,
     Xian 710069, China\\
     ${}^f$ Shaanxi Key Laboratory for Theoretical Physics Frontiers,  Xian 710069, China\\
     ${}^g$ School of Physics, Northwest University,  Xian 710069, China
\end{center}}

\newcommand{\Accepted}[1]{\begin{center}
   {\large \sf #1}\\ \vspace{1mm}{\small \sf Accepted for Publication}
   \end{center}}

\preprint
\thispagestyle{empty}
\bigskip\bigskip\bigskip

\Title{ Exact solution of the $sp(4)$ integrable spin chain with generic boundaries} \Author

\Address
\vspace{1cm}

\begin{abstract}
The off-diagonal Bethe ansatz method is generalized to the integrable model associated with the $sp(4)$  (or $C_2$) Lie algebra.
By using the fusion technique, we obtain the complete  operator product identities among the fused transfer matrices.
These relations, together with some asymptotic behaviors and values of the transfer matrices at certain  points, enable us
to determine the eigenvalues of the transfer matrices completely.  For the periodic boundary condition case, we recover the same $T-Q$ relations  obtained via conventional Bethe ansatz methods previously,
while for the off-diagonal boundary condition case, the eigenvalues are given in terms of  inhomogeneous $T-Q$ relations,
which could not be obtained by the conventional Bethe ansatz methods.
The method developed in this paper can be directly generalized to generic $sp(2n)$ (i.e., $C_n$) integrable model.

\vspace{1truecm} \noindent {\it PACS:} 75.10.Pq, 02.30.Ik, 71.10.Pm

\end{abstract}
\newpage

\section{Introduction}
\label{intro} \setcounter{equation}{0}

Quantum integrable models play important roles in a variety of
fields such as quantum field theory, condensed matter physics and
statistical physics, because they can provide solid benchmarks for
understanding the many-body effects and new physical concepts in
corresponding universal classes \cite{1,2,Alc87,Skl88,Kor93}.

Solving  integrable models without
$U(1)$ symmetry had been an interesting issue and attracted a lot of attentions in the past several decades \cite{Cao03,Nepo02,Bas1,Bas2,Fra08,Bas3,Bas31,Nep13,Bel13,Bel15,Bel15-1,Ava15}.
Recently, a generic method (the off-diagonal Bethe ansatz
(ODBA)) for solving the integrable models with or without obvious
reference states was proposed \cite{Cao1}. With the help of
the proposed inhomogeneous $T-Q$ relations, several typical models without
$U(1)$ symmetry were solved exactly \cite{Cao2}. Based on the eigenvalues obtained via ODBA, the corresponding Bethe-type eigenstates
were also retrieved \cite{Zha13,Hao14}.  The nested ODBA was first proposed to study the quantum spin chain model
related to $A_n$  algebra \cite{Cao14JHEP143}. However, its generalization to integrable models associated with other high-rank Lie algebras such as $B_n$, $C_n$ and $D_n$ are still missing.
In this paper, we generalize the nested ODBA method to $C_2$ spin chain model with both periodic and off-diagonal open boundary conditions.
This method can also be applied to generic $sp(2n)$  quantum integrable spin chains.

The paper is organized as follows. In section 2, we study the $C_2$ model with periodic boundary condition.
The closed functional relations to determine the eigenvalues of the transfer matrix are obtained by the fusion technique. The exact solutions we derived
coincide exactly with those obtained by the conventional Bethe Ansatz methods \cite{NYRes,Bn}.
The exact spectrum of the $C_2$ model with off-diagonal open boundary conditions, which could not be solved via either algebraic or analytic Bethe Ansatz, is given in terms of an inhomogeneous $T-Q$ relation in section 3.
Section 4 is attributed to the concluding remarks.

\section{Periodic $sp(4)$-invariant spin chain}
\label{c2} \setcounter{equation}{0}

\subsection{The system}

Let  ${\rm V}$ denote a $4$-dimensional linear space with an orthonormal basis $\{|i\rangle|i=1,\cdots,4\}$  which endows the fundamental
representation of the $C_2$ algebra.  The $sp(4)$-invariant $R$-matrix $R(u)\in {\rm End}({\rm V}\otimes
{\rm V})$ is  given by its matrix elements \cite{Bn}
\begin{eqnarray}
R_{1\,2}(u)^{ij}_{kl} = u(u+3)\delta_{ik}\delta_{jl}+
(u+3)\delta_{il}\delta_{jk}-u\xi_i\xi_k
\delta_{j\bar{i}}\delta_{k\bar{l}}, \label{rm}
\end{eqnarray}
where  the index $\bar{i}$ is defined by  $i+\bar{i}=5$, $\xi_i=1$
if $i\in \{1, 2\}$ and $\xi_i=-1$ if $i\in \{3,4\}$. Let us take
the notations for simplicity
\begin{eqnarray}
&&R_{1\,2}(u)^{ii}_{ii}=a(u)=(u+1)(u+3),\nonumber\\[4pt]
&&R_{1\,2} (u)^{ij}_{ij}=b(u)=u(u+3),\quad i\ne j, \bar{j},\nonumber\\[4pt]
&&R_{1\,2} (u)^{i\bar{i}}_{\bar{i}i}=c(u)=2u+3,\nonumber\\[4pt]
&&\xi_i\,\xi_j R_{1\,2} (u)^{i\bar{i}}_{j\bar{j}}=d(u)=-u,\quad i\ne j, \bar{j},\nonumber\\[4pt]
&&R_{1\,2} (u)^{i\bar{i}}_{i\bar{i}}=e(u)=u(u+2),\nonumber\\[4pt]
&&R_{1\,2} (u)^{ij}_{j{i}}=g(u)=u+3, \quad i\ne j, \bar{j}.
 \label{2}
\end{eqnarray}
The $R$-matrix (\ref{rm}) enjoys  the  properties:
\begin{eqnarray}
{\rm regularity}&:&R _{12}(0)=\rho_1(0)^{\frac{1}{2}}{\cal P}_{12},\nonumber\\[4pt]
{\rm unitarity}&:&R _{12}(u)R _{21}(-u)=\rho_1(u),\nonumber\\[4pt]
{\rm crossing-unitarity}&:&R _{12}(u)^{t_1}R_{21}(-u-6)^{t_1}=\rho_1(u+3),\nonumber
\end{eqnarray}
where $\rho_1(u)=a(u)a(-u)$, ${\cal P}$ is the permutation operator with the elements ${\cal P}^{ij}_{kl}=\delta_{il}\delta_{jk}$, and $t_i$
denotes the transposition in $i$-th space, $R _{21}(u)={\cal
P}_{12}R _{12}(u){\cal P}_{12}$.  Here and below we adopt the standard notation: for any
matrix $A\in {\rm End}({\rm V})$, $A_j$ is an embedding operator
in the tensor space ${\rm V}\otimes {\rm V}\otimes\cdots$,
which acts as $A$ on the $j$-th space and as an identity on the
other factor spaces; $R_{ij}(u)$ is an embedding operator of
$R$-matrix in the tensor space, which acts as an identity on the
factor spaces except for the $i$-th and $j$-th ones. The $R$-matrix satisfies the quantum Yang-Baxter
equation (QYBE)
\begin{eqnarray}
R _{12}(u-v)R _{13}(u)R_{23}(v)=R_{23}(v)R _{13}(u)R _{12}(u-v).\label{YBE}
\end{eqnarray}

Let us introduce the ``row-to-row"  (or one-row ) monodromy matrix
$T(u)$, which is a $4\times 4$ matrix with operator-valued elements
acting on ${\rm V}^{\otimes N}$,
\begin{eqnarray}
T_0 (u)=R _{01}(u-\theta_1)R _{02}(u-\theta_2)\cdots R _{0N}(u-\theta_N),\label{T1}
\end{eqnarray}
where $\{\theta_j|j=1,\cdots,N\}$ are arbitrary free complex parameters
usually called as inhomogeneous parameters. The transfer matrix $t_p(u)$ of the
associated spin chain with the periodic boundary condition is given by \cite{Kor93}
\begin{eqnarray}
t_p(u)=tr_0T_0 (u).
\end{eqnarray}

The QYBE (\ref{YBE}) of the $R$-matrix implies that one-row monodromy matrix $T(u)$ satisfies the Yang-Baxter relation
\begin{eqnarray}
R _{12}(u-v)T_1 (u) T_2 (v) =T_2 (v)T_1 (u)R_{12}(u-v).
\label{ybe}
\end{eqnarray}
From the above relation, one can prove that the transfer matrices with different spectral parameters
commute with each other, $[t_p(u), t_p(v)]=0$. Then $t_p(u)$ serves as the
generating functional of the conserved quantities, which ensures the
integrability of the $sp(4)$-invariant  spin chain with the periodic boundary condition described by
the  Hamiltonian
\begin{eqnarray}
H_p= \frac{\partial \ln t_p(u)}{\partial
u}|_{u=0,\{\theta_j\}=0}=\sum^{N}_{k=1}H_{k k+1}, \label{Ham-closed}
\end{eqnarray}
where $H_{kk+1}={\cal P}_{k k+1}R_{kk+1}'(u)|_{u=0}$. The periodic boundary condition implies $H_{N\,N+1}=H_{N\,1}$.

\subsection{Fusion}

The $R$-matrix (\ref{rm}) may degenerate to projection operators at some special points of $u$,
which makes it possible for us to do fusion \cite{Kar79-1, Kar79-10, Kar79-2, Kar79-3, Mez92, Zho96}. For example, if $u=-3$, we have
\bea
R _{12}(-3)=  P^{(1) }_{12}\times S_1.\label{Int-R1}\eea
Here $P^{ (1) }_{12}$  is a one-dimensional projection operator with the form
\bea
P^{ (1) }_{12}=|\psi_0\rangle  \langle \psi_0|, \label{1-project}\eea
where $|\psi_0\rangle =\frac{1}{2}(|14\rangle +|23\rangle -|32\rangle -|41\rangle )$ is a vector in the space ${\rm V}\otimes {\rm V}$ and $S_1$ is a constant matrix (we omit its expression because we do not need it).
If $u=-1$, then
\bea
R _{12}(-1)=  P^{(5) }_{12}\times
S_2.\label{Int-R2}\eea
Here $P^{ (5) }_{12}$ is a five-dimensional projection operator with the form
\bea
P^{ (5) }_{12}=\sum_{i=1}^{5}
|{\psi}^{(5)}_i\rangle  \langle {\psi}^{(5)}_i|, \label{2-project}\eea
where the corresponding vectors are \bea
&&|{\psi}^{(5)}_1\rangle =\frac{1}{\sqrt{2}}(|12\rangle -|21\rangle ), \quad |{\psi}^{(5)}_2\rangle =\frac{1}{\sqrt{2}}(|13\rangle -|31\rangle ),
\nonumber\\[4pt]
&&|{\psi}^{(5)}_3\rangle =\frac{1}{{2}}(|14\rangle -|41\rangle +|32\rangle -|23\rangle ), \quad
|{\psi}^{(5)}_4\rangle =\frac{1}{\sqrt{2}}(|24\rangle -|42\rangle ),\nonumber\\[4pt]
&& |{\psi}^{(5)}_5\rangle =\frac{1}{\sqrt{2}}(|34\rangle -|43\rangle ),\nonumber \eea
and $S_2$ is a constant matrix.

From the QYBE (\ref{YBE}), the one-dimensional fusion associated with the projector (\ref{1-project})  leads to
\bea
&&P^{ (1) }_{12}R _{23}(u)R _{13}(u-3)P^{ (1) }_{12}=a(u)e(u-3)P^{ (1) }_{12},\nonumber\\[4pt]
\label{fu-1}
&&P^{ (1) }_{21}R _{32}(u)R _{31}(u-3)P^{ (1) }_{21}=a(u)e(u-3)P^{ (1) }_{21}. \label{hhgg-1}
\eea
From the five-dimensional fusion associated with the projector (\ref{2-project}), we obtain a new fused $\bar R$-matrix
\bea
&&{\bar R}_{\langle 12\rangle\, 3}(u)=[(u+\frac{3}{2})\tilde{\rho}_0(u+\frac{1}{2})]^{-1}  P^{ (5) }_{12}R _{23}(u+\frac{1}{2})R _{13}(u-\frac{1}{2})P^{ (5) }_{12},\nonumber\\[4pt]
&&{\bar R}_{3\,\langle 12\rangle }(u)=[(u+\frac{3}{2})\tilde{\rho}_0(u+\frac{1}{2})]^{-1}  P^{ (5) }_{21}R _{32}(u+\frac{1}{2})R _{31}(u-\frac{1}{2})P^{ (5) }_{21},\label{hhgg-2}
\eea
where $\tilde{\rho}_0(u)=(u-1)(u+3)$. For simplicity, let us denote the resulting five-dimensional fusion space by $\bar{{\rm V}}_{\bar 1}$ which is spanned by $\{|\psi^{(5)}_i\rangle|i=1,\ldots,5\}$.  It is easy to check that the matrix elements of the fused $R$-matrix $\bar{R}_{\bar{1}\,3}(u)\equiv {\bar R}_{\langle 12\rangle\, 3}(u)$  (or $\bar{R}_{3\, \bar{1}}(u)\equiv {\bar R}_{3,\,\langle 12\rangle}(u)$), as functions of $u$, are degree-one polynomials of $u$.  Moreover, we have
 \bea &&\bar R_{{\bar 1}2}(u) \bar R_{2{\bar 1}}(-u)=-(u+\frac{5}{2})(u-\frac{5}{2}), \no\\[4pt]
&&\bar R_{{\bar 1}2}(u)^{t_1} \bar R_{2{\bar 1}}(-u-6)^{t_1}=-(u+\frac{1}{2})(u+\frac{11}{2}), \no\\[4pt]
&&\bar R_{{\bar 1}2}(u-v) \bar R_{{\bar 1}3}(u) R_{{2}3}(v)=R_{{2}3}(v)\bar R_{{\bar 1}3}(u)  \bar R_{{\bar 1}2}(u-v). \eea
At the point of $u=-\frac{5}{2}$, the fused $\bar R$-matrix reduces to a four-dimensional projector
\bea \bar R_{\bar{1}2}(-\frac{5}{2})= P^{(4) }_{{\bar 1}2}\times
\bar{S},\label{Fusion-5-4}\eea
where $\bar{S}$ is a constant matrix and the four-dimensional projector $P^{(4) }_{{\bar 1}2}$ takes the form of
\bea
P^{(4) }_{{\bar 1}2}=\sum_{i=1}^{4} |{\psi}^{(4)}_i\rangle  \langle {\psi}^{(4)}_i|, \label{cjjc}\eea
with the corresponding vectors as
\bea
&&|{\psi}^{(4)}_1\rangle =\frac{1}{\sqrt{5}}(\sqrt{2}|\psi^{(5)}_1\rangle\otimes|3\rangle -\sqrt{2}|\psi^{(5)}_2\rangle\otimes|2\rangle -|\psi^{(5)}_3\rangle\otimes|1\rangle ),\nonumber\\[4pt]
&&|{\psi}^{(4)}_2\rangle =\frac{1}{\sqrt{5}}(-\sqrt{2}|\psi^{(5)}_1\rangle\otimes|4\rangle -\sqrt{2}|\psi^{(5)}_4\rangle\otimes|1\rangle +|\psi^{(5)}_3\rangle\otimes|2\rangle ),\nonumber\\[4pt]
&&|{\psi}^{(4)}_3\rangle =\frac{1}{\sqrt{5}}(-\sqrt{2}|\psi^{(5)}_2\rangle\otimes|4\rangle -\sqrt{2}|\psi^{(5)}_5\rangle\otimes|1\rangle +|\psi^{(5)}_3\rangle\otimes|3\rangle ),\nonumber\\[4pt]
&&|{\psi}^{(4)}_4\rangle =\frac{1}{\sqrt{5}}(-\sqrt{2}|\psi^{(5)}_4\rangle\otimes|3\rangle +\sqrt{2}|\psi^{(5)}_5\rangle\otimes|2\rangle -|\psi^{(5)}_3\rangle\otimes|4\rangle ).\no
\eea

The property (\ref{Fusion-5-4}) allow us to do fusion with $P^{(4) }_{{\bar 1}2}$ again. The results read
\bea &&
R_{\langle {\bar 1}2\rangle \,3} (u)
=(u+5)^{-1}P^{(4) }_{{\bar 1}2}R_{23}(u+2)\bar R_{{\bar 1}3}(u-\frac{1}{2})P^{(4)}_{{\bar 1}2},\nonumber\\[4pt]
&&R_{3\,\langle {\bar 1}2\rangle} (u)=(u+5)^{-1}
P^{(4) }_{2\bar 1}R _{32}(u+2)\bar R_{3\bar 1}(u-\frac{1}{2})P^{(4)}_{2\bar 1}. \label{fu-2} \eea
After taking the correspondence
\bea
|\psi^{(4)}_i\rangle\longrightarrow |i\rangle,\quad i=1,\ldots,4, \label{Correspondence}
\eea
we get the key relations
\bea
R_{\langle {\bar 1}2\rangle \,3} (u)=R_{1\,3}(u),\quad  R_{3\,\langle {\bar 1}2\rangle} (u)=R_{3\,1}(u),\label{Fusion4-1}
\eea
where the $R$-matrices $R_{1\,3}(u)$ and $R_{3\,1}(u)$ are given by (\ref{rm}).


\subsection{$T-Q$ relations}

From the fused $\bar R$-matrix, we can define the fused monodromy matrix
\begin{eqnarray}
\bar T_{\bar 0}(u)=\bar R_{\bar 01}(u-\theta_1)\bar R_{\bar 02}(u-\theta_2)\cdots \bar R_{\bar 0N}(u-\theta_N), \label{T3}
\end{eqnarray}
which is a $5\times 5$ matrix with  operator-valued elements
acting on ${\rm V}^{\otimes N}$. The fused $\bar R$-matrix and the fused monodromy matrix $\bar T(u)$ satisfy the Yang-Baxter relation
\begin{eqnarray}
\bar R_{\bar 12} (u-v) \bar T_{\bar 1}(u) T_2(v)=  T_2(v) \bar T_{\bar 1}(u) \bar R_{\bar 12} (u-v). \label{yybb2}
\end{eqnarray}
The fused transfer matrix is given by
\bea
 \bar{t}_p(u)=tr_{\bar 0} \bar T_{\bar 0}(u).
\eea

Using fusion relations (\ref{hhgg-1}), (\ref{hhgg-2}) and (\ref{fu-2}), we have \bea
&&P^{ (1) }_{21}T_1 (u)T_2 (u-3)P^{(1) }_{21}=\prod_{i=1}^N a(u-\theta_i)e(u-\theta_i-3)P^{(1)
}_{21},\label{fut-1}\\
&&P^{ (5) }_{21}T_1 (u)T_2(u-1)P^{ (5) }_{21}=\prod_{i=1}^N
(u-\theta_i+1)\tilde{\rho}_0(u-\theta_i)\bar T_{\langle 12\rangle}(u-\frac{1}{{2}}),\label{fut-2}\\
&&P^{(4) }_{{\bar 1}2} T_2 (u)\bar T_{{\bar 1}}(u-\frac{5}{{2}})P^{(4)
}_{{\bar 1}2}=\prod_{i=1}^N (u-\theta_i+3)T_{\langle {\bar 1}2\rangle }(u-2).\label{fut-6}
 \eea
Following the method developed in \cite{Cao14JHEP143} and using the identity (\ref{Fusion4-1}), we can show that the following identities hold
\bea &&T_1 (\theta_j)T_2 (\theta_j-3)=P^{ (1)
}_{21}T_1(\theta_j)T_2 (\theta_j-3),\label{fui-1}\\[4pt]
&&T_1 (\theta_j)T_2 (\theta_j-1)=P^{ (5)
}_{21}T_1(\theta_j)T_2 (\theta_j-1),\label{fui-2}\\[4pt]
&&T_2 (\theta_j)\bar T_{ \bar 1}(\theta_j-\frac{5}{{2}})=P^{(4)
}_{{\bar 1}2} T_2 (\theta_j)\bar T_{ \bar 1}(\theta_j-\frac{5}{{2}}).\label{fui-5}\eea
Considering the relations (\ref{Fusion4-1}) and  (\ref{fut-1})-(\ref{fui-5}), we obtain the operator identities among the fused transfer matrices as
\bea && t_p(\theta_j)t_p (\theta_j-3)=\prod_{i=1}^N
a(\theta_j-\theta_i)e(\theta_j-\theta_i-3),\label{futp-1} \\[4pt]
&& t_p (\theta_j)t_p (\theta_j-1)= \prod_{i=1}^N
(\theta_j-\theta_i+1)\tilde{\rho}_0(\theta_j-\theta_i)\bar t_p(\theta_j-\frac{1}{{2}}),\label{futp-2} \\[4pt]
&& t_p (\theta_j)\bar t_p(\theta_j-\frac{5}{{2}})=\prod_{i=1}^N
(\theta_j-\theta_i+3)t_p(\theta_j-2). \label{futp-6} \eea

The commutativity  of the transfer matrices $t_p(u)$ and $\bar{t}_p(u)$  with different spectral parameters implies that they have common eigenstates (namely, the common eigenstates do not depend on the spectral parameter $u$). Let us denote the eigenvalues of the transfer matrices $t_p(u)$ and $\bar t_p(u)$ as
$\Lambda_p(u)$ and $\bar \Lambda_p(u)$, respectively. From the identities (\ref{futp-1})-(\ref{futp-6}), we have
\bea && \Lambda_p (\theta_j)\Lambda_p(\theta_j-3)=\prod_{i=1}^N
a(\theta_j-\theta_i)e(\theta_j-\theta_i-3),\label{futpl-1} \\[4pt]
&& \Lambda_p(\theta_j)\Lambda_p (\theta_j-1)=
\prod_{i=1}^N(\theta_j-\theta_i+1)
\tilde{\rho}_0(\theta_j-\theta_i)\bar \Lambda_p(\theta_j-\frac{1}{{2}}),\label{futpl-2} \\[4pt]
&& \Lambda_p (\theta_j)\bar
\Lambda_p(\theta_j-\frac{5}{{2}})=\prod_{i=1}^N
(\theta_j-\theta_i+3)\Lambda_p(\theta_j-2).\label{futpl-6} \eea
The eigenvalue $\Lambda_p(u)$ of the transfer matrix $t_p(u)$ is a degree $2N$
polynomial of $u$, which can be completely
determined by $2N + 1$ conditions. Besides the functional
relations (\ref{futpl-1})-(\ref{futpl-6}), we still need one more
condition which can be obtained by analyzing the asymptotic
behavior of $t_p(u)$. From the definition, the asymptotic behavior
of $t_p(u)$ can be calculated as \bea t_p(u)|_{u\rightarrow
\infty}= 4u^{2N}\times {\rm id}+\cdots,\nonumber \eea which leads
to \bea \Lambda_p(u)|_{u\rightarrow  \infty}= 4u^{2N}+\cdots.
\label{tpjixiandian} \eea The eigenvalue $\bar \Lambda_p(u)$ of
the fused transfer matrix $\bar t_p(u)$ is a  degree $N$ polynomial of $u$, which can be completely determined by the
functional relations (\ref{futpl-1})-(\ref{futpl-6}) and the
asymptotic behavior of $\bar t_p(u)$ given by  \bea \bar
t_p(u)|_{u\rightarrow \infty}=5u^{N}\times {\rm
id}+\cdots,\nonumber\eea
or \bea \bar
\Lambda_p(u)|_{u\rightarrow
\infty}=5u^{N}+\cdots.\label{Asym-2}\eea Thus the $3N+2$ relations
(\ref{futpl-1})-(\ref{Asym-2}) completely determine the
eigenvalues $\Lambda_p(u)$ and $\bar \Lambda_p(u)$, which are
given in terms of the homogeneous $T-Q$ relations:
 \bea &&\Lambda_p (u)=Z^{(p)}_{1} (u)+
Z^{(p)}_{2}(u)+Z^{(p)}_{3}(u)+Z^{(p)}_{4}(u),\label{ep-1}\\[4pt]
&&\bar \Lambda_p(u)= \prod_{i=1}^N[(u-\theta_i+\frac32)
\tilde{\rho}_0(u-\theta_i+\frac12)]^{-1}\left\{
Z^{(p)}_{1}(u+\frac12)[
Z^{(p)}_{2}(u-\frac12)+Z^{(p)}_{3}(u-\frac12)\rt.\no\\[4pt]
&&\qquad\quad \lt.+Z^{(p)}_{4}(u-\frac12)]+[Z^{(p)}_{2}(u+\frac12)+Z^{(p)}_{3}(u+\frac12)]Z^{(p)}_{4}(u-\frac12)\right\},\label{ep-3}\eea
where \bea &&Z^{(p)}_{1}(u)=\prod_{j=1}^N
a(u-\theta_j)\frac{Q_p^{(1)}(u-1)}{Q_p^{(1)}(u)},\no\\[4pt]
&&Z^{(p)}_{2}(u)=\prod_{j=1}^N
b(u-\theta_j)\frac{Q_p^{(1)}(u+1)Q_p^{(2)}(u-\frac{3}{2})}{Q_p^{(1)}(u)Q_p^{(2)}(u+\frac{1}{2})},\no\\[4pt]
&&Z^{(p)}_{3}(u)=\prod_{j=1}^N
b(u-\theta_j)\frac{Q_p^{(1)}(u+1)Q_p^{(2)}(u+\frac{5}{2})}{Q_p^{(1)}(u+2)Q_p^{(2)}(u+\frac{1}{2})},\no\\[4pt]
&&Z^{(p)}_{4}(u)=\prod_{j=1}^N e(u-\theta_j)
\frac{Q_p^{(1)}(u+3)}{Q_p^{(1)}(u+2)},\no\\[4pt]
&&Q^{(m)}_p(u)=\prod_{k=1}^{L_m}(u-\mu_k^{(m)}+\frac{m}{2}), \quad m=1, 2.\eea
Because the eigenvalues $\Lambda_p(u)$ and $\bar \Lambda_p(u)$ must be polynomials of $u$, the residues of
right hand sides of equations (\ref{ep-1})-(\ref{ep-3}) should be
zero, which gives rise to  the constraints of the Bethe roots
$\{\mu_k^{(m)}\}$, namely,  these parameters should satisfy the Bethe Ansatz equations
\bea &&
\frac{Q^{(1)}_p(\mu_k^{(1)}+\frac{1}{2})Q^{(2)}_p(\mu_k^{(1)}-2)}{Q^{(1)}_p(\mu_k^{(1)}-\frac{3}{2})Q^{(2)}_p(\mu_k^{(1)})}
=-\prod_{j=1}^N\frac{\mu_k^{(1)}+\frac{1}{2}-\theta_j
}{\mu_k^{(1)}-\frac{1}{2}-\theta_j},\quad k=1,\cdots, L_1, \label{BAE-period-1}\\[4pt]
&&\frac{Q^{(2)}_p(\mu_l^{(2)}+1)Q^{(1)}_p(\mu_l^{(2)}-\frac{3}{2})}{Q^{(2)}_p(\mu_l^{(2)}-3)Q^{(1)}_p(\mu_l^{(2)}+\frac{1}{2})}
=-1, \quad l=1,\cdots, L_2. \label{BAE-period-2}\eea We note that
the Bethe ansatz equations obtained from the regularity of
$\Lambda_p(u)$ are  the same as those obtained from the regularity
of $\bar \Lambda_p(u)$. It is easy to check  that $\Lambda_p (u)$
and $\bar \Lambda_p(u)$ satisfy the functional relations
(\ref{futpl-1})-(\ref{futpl-6}) and the asymptotic behaviors
(\ref{tpjixiandian}) and (\ref{Asym-2}). Therefore, we conclude
that $\Lambda_p(u)$ and $\bar \Lambda_p(u)$ are the eigenvalues of
the transfer matrices $t_p(u)$ and $\bar t_p(u)$, respectively. It
is remarked that the $T-Q$ relation (\ref{ep-1}) and the
associated Bethe ansatz equations
(\ref{BAE-period-1})-(\ref{BAE-period-2})(after taking the homogeneous limit $\{\theta_j\to0|j=1,2,\cdots,N\}$) coincide with those
obtained previously via conventional Bethe ansatz
methods\cite{NYRes,Bn} .

The eigenvalues of the Hamiltonian (\ref{Ham-closed}) then can  be expressed in terms of the Bethe roots as
\begin{eqnarray}
E_p= \frac{\partial \ln \Lambda_p(u)}{\partial
u}|_{u=0,\{\theta_j\}=0}.
\end{eqnarray}

\section{Off-diagonal integrable open boundary case}
\setcounter{equation}{0}

\subsection{Open chain}
Integrable open chain can be constructed as follows \cite{Alc87,Skl88}.
Let us introduce a pair of $K$-matrices $K^-(u)$ and $K^+(u)$. The former satisfies the reflection equation (RE)
\begin{equation}
 R _{12}(u-v){K^{-}_{  1}}(u)R _{21}(u+v) {K^{-}_{2}}(v)=
 {K^{-}_{2}}(v)R _{12}(u+v){K^{-}_{1}}(u)R _{21}(u-v),
 \label{r1}
 \end{equation}
and the latter  satisfies the dual RE
\begin{eqnarray}
&&R _{12}(-u+v){K^{ +}_{1}}(u)R _{21}
 (-u-v-6){K^{ +}_{2}}(v)\nonumber\\
&&\qquad\qquad={K^{ +}_{2}}(v)R _{12}(-u-v-6) {K^{ +}_{1}}(u)R _{21}(-u+v).
 \label{r2}
 \end{eqnarray}
For open spin-chains, instead of the standard
``row-to-row" monodromy matrix $T(u)$ (\ref{T1}), one needs to
consider  the ``double-row" monodromy matrix  as follows.
Besides the monodromy matrix $T_0 (u)$ given by (\ref{T1}), we also need a crossed monodromy matrix
\begin{eqnarray}
\hat{T}_0 (u)=R_{N0}(u+\theta_N)\cdots R_{20}(u+\theta_{2}) R_{10}(u+\theta_1),\label{Tt11}
\end{eqnarray}
which satisfies the Yang-Baxter relation
\begin{eqnarray}
R_{ 12} (u-v) \hat T_{1}(u) \hat T_2(v)=\hat  T_2(v) \hat T_{ 1}(u) R_{12} (u-v).\label{haishi0}
\end{eqnarray}
The transfer matrix $t(u)$ is defined as \cite{Skl88}
\begin{equation}
t(u)= tr_0 \{K_0^{ +}(u)T_0 (u) K^{ -}_0(u)\hat{T}_0 (u)\}.\label{tru}
\end{equation}
From the Yang-Baxter relation, reflection equation and dual reflection equation, one can
prove that the transfer matrices with different spectral parameters
commute with each other, $[t(u), t(v)]=0$. Therefore, $t(u)$ serves
as the generating function of all the conserved quantities of the
system. The associated quantum spin chain with integrable boundary interactions is given by  the Hamiltonian
\begin{eqnarray}
H&=&\frac{1}{2}\,\frac{\partial \ln t(u)}{\partial
u}|_{u=0,\{\theta_j\}=0} \nonumber \\[4pt]
&=& \sum^{N-1}_{k=1}H_{k k+1}+\frac{{K^{-}_1}'(0)}{2\zeta}+\frac{
tr_0 \{K^{+}_0(0)H_{N0}\}}{tr_0 K^{+}_0(0)}. \label{hh}
\end{eqnarray}

In this paper, we consider the open chain  with off-diagonal $K$-matrix  $K^-(u)$ \cite{8Mel054,8Mel055,8Mel056}
\bea
K^{-}(u)=\zeta+M
u, \quad
M=\left(\begin{array}{cccc}-1 &0&c_{1}&0\\[6pt]
0&-1 &0&c_{1}\\[6pt]
c_{2}&0 &1 &0\\[6pt]
0&c_{2}&0&1 \end{array}\right),\label{K-matrix-1}\eea
while the dual reflection matrix $K^+(u)$ is
\begin{equation}
K^{ +}(u)=K^{ -}(-u-3)|_{\zeta,c_i\rightarrow
\tilde{\zeta},\tilde{c}_i }. \label{ksk}
\end{equation}
Here $\zeta$, $c_1$,  $c_2 $ and $\tilde{\zeta}$, $\tilde{c}_1$,
$\tilde{c}_2 $ are some generic  parameters describing
the boundary fields applied on the end sites. It is easily to check that
$[K^-(u),\,K^+(v)]\neq 0$, which implies that the
$K^{\pm}(u)$ matrices  cannot be diagonalized simultaneously. In this case, it is quite hard to derive solutions via
conventional Bethe Ansatz methods
due to the absence of a proper reference state.
We will generalize the method developed in
Section 2 to get eigenvalues of the transfer matrix $t(u)$
(\ref{tru}) specified by the $K$-matrices (\ref{K-matrix-1}) and
(\ref{ksk}) in the following subsections.

\subsection{Fusion of the reflection matrices}

In order to obtain closed operator identities, we study first fusions of the reflection matrices. The one-dimensional fusion for the reflection matrices gives
\bea && P_{21}^{ (1)}K_{1}^{-}(u)R_{21} (2u-3)K_{2}^{-}(u-3)P_{12}^{ (1)} =(u-1)(u-3)h(u)P_{12}^{ (1)},\label{fk-1}\\[4pt]
&& P_{12}^{ (1)}K_{2}^{+}(u-3)R_{12} (-2u-3)K_{1}^{+}(u)P_{21}^{ (1)} =(u+1)(u+3)\tilde{h}(u)P_{21}^{ (1)},\label{fk-2}\eea where
\bea &&
h(u)={4}[(1+c_1c_2)u^2-\zeta^2],\no\\[4pt]
&&
\tilde{h}(u)={4}[(1+\tilde{c}_1\tilde{c}_2)u^2-\tilde{\zeta}^2].\no
 \eea
From the five-dimensional fusion, we obtain a new fused reflection matrices $\bar K$ as
\bea
&&\bar K_{\langle 12\rangle}^{-}(u)= [(2u\hspace{-0.12truecm}-\hspace{-0.12truecm}1)(2u\hspace{-0.12truecm}+\hspace{-0.12truecm}3)]^{-1}
P_{21}^{ (5)}K_{1}^{-}(u\hspace{-0.12truecm}+\hspace{-0.12truecm}\frac{1}{{2}})
R_{21} (2u)K_{2}^{-}(u\hspace{-0.12truecm}-\hspace{-0.12truecm}\frac{1}{{2}})
P_{12}^{ (5)},\label{fu-1qwe3}\\[6pt]
&&\bar K_{\langle 12\rangle }^{+}(u) = [(2u\hspace{-0.12truecm}+\hspace{-0.12truecm}3)(2u\hspace{-0.12truecm}+\hspace{-0.12truecm}7)]^{-1} P_{12}^{ (5)}K_{2}^{+}(u\hspace{-0.12truecm}-\hspace{-0.12truecm}\frac{1}{{2}})R_{12} (\hspace{-0.12truecm}-\hspace{-0.12truecm}2u\hspace{-0.12truecm}-\hspace{-0.12truecm}6)K_{1}^{+}(u\hspace{-0.12truecm}+\hspace{-0.12truecm}\frac{1}{{2}})
P_{21}^{ (5)},\label{fu-3}\eea
where the projector $P_{12}^{ (5)}$ is given by (\ref{2-project}). Due to the dimension of the projected space $\bar{{\rm V}}$ is 5,  the corresponding $\bar K_{\bar 1}^{\pm}(u)$ are both $5\times 5$ matrices and their elements are all  degree two polynomials of $u$.
The fused $\bar R$-matrix and the fused reflection matrix $\bar K^{\pm}(u)$ satisfy the reflection equations
\begin{eqnarray}
&& \bar R _{\bar 12}(u-v)\bar K^{-}_{\bar  1}(u) \bar R _{2\bar 1}(u+v) K^{-}_{2}(v)=K^{-}_{2}(v)
\bar R_{\bar 12}(u+v) \bar K^{-}_{\bar 1}(u) \bar R _{2\bar 1}(u-v), \\[4pt]
&& \bar R _{\bar 12}(-u+v) \bar K^{+}_{\bar 1}(u) \bar R _{2\bar 1}
 (-u-v-6) K^{ +}_{2}(v) \nonumber\\[4pt]
&&\qquad =K^{ +}_{2}(v) \bar R _{\bar 12}(-u-v-6) \bar K^{ +}_{\bar 1}(u)\bar R _{2\bar 1}(-u+v). \label{haishi8}
\end{eqnarray}

Now let us turn to the fusion between the reflection matrices $K^{\pm}(u)$ and $\bar K^{\pm}(u)$ by the four-dimensional projector $ P_{{\bar 1}2}^{(4)}$ given by (\ref{cjjc}), which gives
\bea && K^{-}_{\langle {\bar 1}2\rangle }(u)={4}[(2u-1)h(u+2)]^{-1}
P_{{\bar1}2}^{(4)}K_{2}^{-}(u+2)\bar R_{\bar 12}(2u+\frac{3}{{2}})\bar K_{\bar 1}^{-}(u-\frac{1}{{2}})P_{2{\bar 1}}^{(4)},\no \\[4pt]
&& K^{+}_{\langle {\bar 1}2\rangle }(u)=-{2}[(u+5)\tilde{h}(u+2)]^{-1}
P_{2{\bar 1}}^{(4)}\bar K_{\bar 1}^{+}(u-\frac{1}{{2}})\bar R_{2{\bar 1}}(-2u-\frac{15}{{2}})K_{2}^{+}(u+2)P_{{\bar 1}2}^{(4)}.\no \eea
Both $K^{\pm}_{\langle {\bar 1}2\rangle }(u)$ are $4\times 4$ matrices, whose matrix elements are degree one polynomials of $u$. Moreover, keeping the correspondence (\ref{Correspondence}) in mind, we have
\bea
K^{\pm}_{\langle {\bar 1}2\rangle }(u)\equiv K^{\pm}(u),\label{Correspondence-2}
\eea
where the $K$-matrices $K^{\pm}(u)$ are given by (\ref{K-matrix-1}) and (\ref{ksk}).

\subsection{Operator product identities}

For the open boundary case, besides the fused monodromy matrix $\bar T_{\bar 0} (u)$ given by (\ref{T3}), we also need the reflecting fused monodromy matrix $\hat{\bar T}(u)$ given by
\begin{eqnarray}
\hat{\bar T}_{\bar 0}(u)=\bar R_{N{\bar 0}}(u+\theta_N)\cdots \bar R_{2{\bar 0}}(u+\theta_{2}) \bar R_{1{\bar 0}}(u+\theta_1),\label{Tt4}
\end{eqnarray}
where the dimension of auxiliary space $\bar{{\rm V}}$ is $5$ and the quantum space keeps unchanged.
The matrix $\hat{\bar T}_{\bar 0}$ satisfies the Yang-Baxter relation
\begin{eqnarray}
\bar R_{\bar 12} (u-v) \hat {\bar T}_{\bar 1}(u) \hat T_2(v)=  \hat T_2(v) \hat {\bar T}_{\bar 1}(u) \bar R_{\bar 12} (u-v). \label{haishi}
\end{eqnarray}
The fused transfer matrix $\bar t(u)$ is
\bea
\bar t(u)= tr_{\bar 0}\{\bar K_{\bar 0}^{+}(u)  \bar T_{\bar 0}(u)    \bar K_{\bar 0}^{-}(u)\hat{\bar T}_{\bar 0}(u)\}.\label{transfer-2}
\eea
From the definitions (\ref{tru}) and (\ref{transfer-2}),  the transfer matrix $t(u)$ (resp. $\bar t(u)$),
as a function of $u$, is a polynomial of degree $4N+2$ (resp. a polynomial with degree of $2N+4$). In order to determine
the eigenvalues of the transfer matrices $t(u)$ and $\bar t(u)$, one needs  at least $6N+8$ conditions.

With the fusion relations (\ref{hhgg-1}), (\ref{hhgg-2}) and (\ref{fu-2}), we have \bea
&&P^{ (1) }_{12}\hat{T}_1 (u)\hat{T}_2(u-3)P^{ (1) }_{12}=\prod_{i=1}^N
a(u+\theta_i)e(u+\theta_i-3)P^{ (1)
}_{12},\label{futt-1}\\
&&P^{ (5) }_{12}\hat{T}_1(u)\hat{T}_2 (u-1)P^{ (5)
}_{12}=\prod_{i=1}^N(u+\theta_i+1)
\tilde{\rho}_0(u+\theta_i)\hat{\bar T}_{\langle {1}2\rangle }(u-\frac{1}{{2}}),\label{futt-2}\\
&&P^{(4) }_{2{\bar 1}}\hat{T}_2(u)\hat{\bar T}_{\bar 1}(u-\frac{5}{2})P^{(4) }_{2{\bar1}}=\prod_{i=1}^N
(u+\theta_i+3)\hat{T}_{\langle{\bar 1}2\rangle}(u-2).\label{futt-7}
 \eea
We can show
that Yang-Baxter relations (\ref{haishi0}) and (\ref{haishi}) at
certain points also give
\bea &&\hat{T}_1 (-\theta_j)\hat{T}_2(-\theta_j-\frac{3}{{2}})=P^{ (1) }_{12}\hat{T}_1(-\theta_j)\hat{T}_2 (-\theta_j-\frac{3}{{2}}),\label{fuii-1}\\[4pt]
&&\hat{T}_1 (-\theta_j)\hat{T}_2(-\theta_j-1)=P^{ (5) }_{12}\hat{T}_1(-\theta_j)\hat{T}_2 (-\theta_j-1),\label{fuii-2}\\[4pt]
&&\hat{T}_2 (-\theta_j)\hat{\bar T}_{\bar 1}(-\theta_j-\frac{5}{2})=P^{(4)
}_{2{\bar 1}}\hat{T}_2 (-\theta_j)\hat{\bar T}_{\bar 1}(-\theta_j-\frac{5}{2}).\label{fuii-7} \eea
Keeping the identity (\ref{Fusion4-1}) in mind and  using the relations (\ref{fut-1})-(\ref{fui-5}), (\ref{Correspondence-2})  and (\ref{futt-1})-(\ref{fuii-7}),
we obtain
\bea && t (\pm\theta_j)t (\pm\theta_j-3)=\frac{1}{
2^{4}} \frac{(\pm\theta_j-1)(\pm\theta_j-3)
(\pm\theta_j+1)(\pm\theta_j+3)}{(\pm\theta_j-\frac{3}{{2}})(\pm\theta_j-\frac{1}{{2}})
(\pm\theta_j+\frac{3}{{2}})(\pm\theta_j+\frac{1}{{2}})}\nonumber\\[4pt]
&&\hspace{5mm}\times
h(\pm\theta_j)\tilde{h}(\pm\theta_j)\prod_{i=1}^N
a(\pm\theta_j-\theta_i)a(\pm\theta_j+\theta_i)e(\pm\theta_j-\theta_i-3)e(\pm\theta_j+\theta_i-3),\label{ftpp-1} \\[4pt]
&& t (\pm\theta_j)t (\pm\theta_j-1)= \frac{
(\pm\theta_j-1)(\pm\theta_j+1)(\pm\theta_j+1)(\pm\theta_j+3)}{(\pm\theta_j-\frac{1}{{2}})(\pm\theta_j+\frac{1}{{2}})
(\pm\theta_j+\frac{3}{{2}})(\pm\theta_j+\frac{5}{{2}})} \no\\[4pt]
&&\hspace{10mm}\times \prod_{i=1}^N
(\pm\theta_j-\theta_i+1)(\pm\theta_j+\theta_i+1)\tilde{\rho}_0(\pm\theta_j-\theta_i)\tilde{\rho}_0(\pm\theta_j+\theta_i){\bar t}(\pm\theta_j-\frac{1}{{2}}), \label{ftpp-2} \\[4pt]
&& t (\pm\theta_j){\bar t}(\pm\theta_j-\frac{5}{{2}})=\frac{1}{
2^{4}}
\frac{(\pm\theta_j-\frac{5}{{2}})(\pm\theta_j+3)}{(\pm\theta_j-1)
(\pm\theta_j+\frac{3}{{2}})}h(\pm\theta_j)\tilde{h}(\pm\theta_j)\nonumber\\[4pt]
&&\hspace{40mm}\times
 \prod_{i=1}^N  (\pm\theta_j-\theta_i+3)(\pm\theta_j+\theta_i+3)  t(\pm\theta_j-2).\label{ftpp-7} \eea
Form the definition, the asymptotic behavior of $t(u)$ can be
calculated as \bea t(u)|_{u\rightarrow \infty} =  -tr\tilde{M}{M}
\times u^{4N+2} \times {\rm id}+\cdots, \label{jixian1} \eea where
$\tilde{M}=M|_{\zeta,c_1,c_2\rightarrow
\tilde{\zeta},\tilde{c}_1,\tilde{c}_2 }$. Direct calculation gives
\bea tr\tilde{M}{M}=4+2c_1\tilde{c}_2+2c_2\tilde{c}_1. \eea
Besides, we also know the values of $t(u)$ at the points of $0$
and $-3$, \bea &&t (0)=  tr \{K^{+}(0)\}\ {\zeta}\
\prod_{i=1}^N\rho_1
(-\theta_i)\,\times{\rm id},\label{t1-1} \\[4pt]
&&t (-3)= tr \{K^{-}(-3)\}\ \tilde{\zeta}\ \prod_{i=1}^N\rho_1
(-\theta_i)\,\times{\rm id}.\label{t1-11} \eea The asymptotic
behavior of $\bar t(u)$ reads \bea \bar t (u)|_{u\rightarrow
\infty}  = tr_{12}P^{ (5) }_{12}(\tilde{M}{M})_1(\tilde{M}{M})_2
P^{ (5) }_{12}
 \times u^{2N+4} \times {\rm id} +\cdots. \label{jixian2}
\eea Direct calculation shows \bea
tr_{12}P^{ (5) }_{12}(\tilde{M}{M})_1(\tilde{M}{M})_2
P^{ (5) }_{12}=(2+c_1\tilde{c}_2+c_2\tilde{c}_1)^2+(1+c_1c_2)(1 +\tilde{c}_1\tilde{c}_2).\no\eea
Using the method developed in \cite{Cao14JHEP143}, we can evaluate the values of $\bar t(u)$ at some special points as follows:
\bea
&&\bar t (0)=\frac{5}{2^4}(1+c_1c_2-4\zeta^2)(1+\tilde{c}_1\tilde{c}_2-4\tilde{\zeta}^2)\,\prod_{i=1}^N
(\frac{5}{2}-\theta_i)(\frac{5}{2}+\theta_i)\times {\rm id}, \label{t1-2}\\[4pt]
&&\bar t (-3)=\frac{5}{2^4}(1+c_1c_2-4\zeta^2)(1+\tilde{c}_1\tilde{c}_2-4\tilde{\zeta}^2)\,\prod_{i=1}^N
(\frac{5}{2}-\theta_i)(\frac{5}{2}+\theta_i)\times {\rm id},\label{t1-22}\\[4pt]
&&\bar t (-\frac{1}{2})=\frac{5}{4}\frac{\zeta\tilde{\zeta}}{\prod_{i=1}^N
(1-\theta_i)(1+\theta_i)}\,  t(-1),\label{t2-1} \\[4pt]
&&\bar t
(-\frac{5}{2})=\frac{5}{4}\frac{\zeta\tilde{\zeta}}{\prod_{i=1}^N
(1-\theta_i)(1+\theta_i)}\,  t(-2).\label{t2-11}
\eea

\subsection{Inhomogeneous $T-Q$ relations}
So far we have obtained the $6N+8$ conditions (\ref{ftpp-1})-(\ref{t2-11}),  which allow us to determine the eigenvalues of the transfer matrices $t(u)$ and $\bar t(u)$.
It is easy to show that the transfer matrix $t(u)$ and its fused one $\bar t(u)$ satisfy the commutation relations
\bea
[t(u),\,t(v)]= [\bar t(u),\,\bar t(v)]=[t(u),\,\bar t(v)]=0.\nonumber
\eea
Let $|\Psi\rangle$  be a common eigenstate of the transfer matrices with the eigenvalues $\Lambda(u)$ and $\bar \Lambda(u)$
\bea
t(u)|\Psi\rangle =\Lambda(u)|\Psi\rangle,\quad \bar t(u)|\Psi\rangle =\bar \Lambda(u)|\Psi\rangle.\nonumber
\eea
From the identities (\ref{ftpp-1})-(\ref{ftpp-7}), we obtain the following closed functional relations
\bea && \Lambda (\pm\theta_j)\Lambda(\pm\theta_j-3)=\frac{1}{ 2^{4}}
\frac{(\pm\theta_j-1)(\pm\theta_j-3)
(\pm\theta_j+1)(\pm\theta_j+3)}{(\pm\theta_j-\frac{3}{{2}})(\pm\theta_j-\frac{1}{{2}})
(\pm\theta_j+\frac{3}{{2}})(\pm\theta_j+\frac{1}{{2}})}\nonumber\\[4pt]
&&\hspace{5mm}\times
h(\pm\theta_j)\tilde{h}(\pm\theta_j)\prod_{i=1}^N
a(\pm\theta_j-\theta_i)a(\pm\theta_j+\theta_i)e(\pm\theta_j-\theta_i-3)e(\pm\theta_j+\theta_i-3),\label{fetpp-1} \\[4pt]
&& \Lambda (\pm\theta_j)\Lambda (\pm\theta_j-1)=
\frac{
(\pm\theta_j-1)(\pm\theta_j+1)(\pm\theta_j+1)(\pm\theta_j+3)}{(\pm\theta_j-\frac{1}{{2}})(\pm\theta_j+\frac{1}{{2}})
(\pm\theta_j+\frac{3}{{2}})(\pm\theta_j+\frac{5}{{2}})}\no\\[4pt]
&&\hspace{10mm}\times \prod_{i=1}^N
(\pm\theta_j-\theta_i+1)(\pm\theta_j+\theta_i+1)\tilde{\rho}_0(\pm\theta_j-\theta_i)\tilde{\rho}_0(\pm\theta_j+\theta_i)\bar  \Lambda(\pm\theta_j-\frac{1}{{2}}), \label{fetpp-2} \\[4pt]
&&\Lambda (\pm\theta_j)\bar \Lambda
(\pm\theta_j-\frac{5}{{2}})=\frac{1}{ 2^{4}}
\frac{(\pm\theta_j-\frac{5}{{2}})(\pm\theta_j+3)}{(\pm\theta_j-1)
(\pm\theta_j+\frac{3}{{2}})}h(\pm\theta_j)\tilde{h}(\pm\theta_j)\nonumber\\[4pt]
&&\hspace{40mm}\times
 \prod_{i=1}^N  (\pm\theta_j-\theta_i+3)(\pm\theta_j+\theta_i+3) \Lambda (\pm\theta_j-2),\label{fetpp-7} \\[4pt]
 &&\Lambda(u)|_{u\rightarrow
\infty} =  -(4+2c_1\tilde{c}_2+2c_2\tilde{c}_1) u^{4N+2} +\cdots,\label{fetpp-8}\\[4pt]
&&\Lambda(0)= 4\zeta\tilde{\zeta}\ \prod_{i=1}^N\rho_1(-\theta_i),\label{fetpp-9} \\[4pt]
&&\Lambda(-3)= 4\zeta\tilde{\zeta}\ \prod_{i=1}^N\rho_1(-\theta_i),\label{fetpp-10}\\[4pt]
&&\bar \Lambda(u)|_{u\rightarrow\infty}  = \left\{(2+c_1\tilde{c}_2+c_2\tilde{c}_1)^2+(1+c_1c_2)(1 +\tilde{c}_1\tilde{c}_2)\right\}u^{2N+4}  +\cdots,\label{fetpp-11}\\[4pt]
&&\bar
\Lambda(0)=\frac{5}{2^4}(1+c_1c_2-4\zeta^2)(1+\tilde{c}_1\tilde{c}_2-4\tilde{\zeta}^2)\prod_{i=1}^N
(\frac{5}{2}-\theta_i)(\frac{5}{2}+\theta_i), \label{fetpp-12}\\[4pt]
&&\bar
\Lambda(-3)=\frac{5}{2^4}(1+c_1c_2-4\zeta^2)(1+\tilde{c}_1\tilde{c}_2-4\tilde{\zeta}^2)\prod_{i=1}^N
(\frac{5}{2}-\theta_i)(\frac{5}{2}+\theta_i),\label{fetpp-13}\\[4pt]
&&\bar
\Lambda(-\frac{1}{2})=\frac{5}{4}\frac{\zeta\tilde{\zeta}}{\prod_{i=1}^N
(1-\theta_i)(1+\theta_i)}\,\Lambda(-1),\label{fetpp-14} \\[4pt]
&&\bar
\Lambda(-\frac{5}{2})=\frac{5}{4}\frac{\zeta\tilde{\zeta}}{\prod_{i=1}^N
(1-\theta_i)(1+\theta_i)} \Lambda(-2).\label{fetpp-15} \eea

The $6N+8$ relations (\ref{fetpp-1})-(\ref{fetpp-15}) enable us completely to determine the eigenvalues $\Lambda(u)$ and $\bar \Lambda(u)$ which are given in terms of some inhomogeneous $T-Q$ relations. For simplicity, we first define some functions:
 \bea &&Z_1(u)=\frac{1}{ 2^{2}} \frac{
(u+1)(u+3)}{(u+\frac{1}{{2}})(u+\frac{3}{{2}}) }\prod_{j=1}^N
a(u-\theta_j)a(u+\theta_j)\frac{Q^{(1)}(u-1)}{Q^{(1)}(u)} {h_1(u)\tilde{h}_1(u)},\no\\[4pt]
&&Z_2(u)=\frac{1}{ 2^{2}} \frac{
u(u+3)}{(u+\frac{1}{{2}})(u+\frac{3}{{2}}) }\prod_{j=1}^N
b(u-\theta_j)b(u+\theta_j)\frac{Q^{(1)}(u+1)Q^{(2)}(u-\frac{3}{{2}})}{Q^{(1)}(u)Q^{(2)}(u+\frac{1}{{2}})}{h_1(u)\tilde{h}_1(u)},\no\\[4pt]
&&Z_3(u)=\frac{1}{ 2^{2}}\frac{
u(u+3)}{(u+\frac{3}{{2}})(u+\frac{5}{{2}}) }\prod_{j=1}^N
b(u-\theta_j)b(u+\theta_j) \nonumber \\[4pt]
&&\qquad\qquad \times \frac{Q^{(1)}(u+1)Q^{(2)}(u+\frac{5}{2})}{Q^{(1)}(u+2)Q^{(2)}(u+\frac{1}{2})}{h_2(u+3)\tilde{h}_2(u+3)},\no\\[4pt]
&&Z_4(u)=\frac{1}{ 2^{2}} \frac{
u(u+2)}{(u+\frac{3}{{2}})(u+\frac{5}{{2}}) }\prod_{j=1}^N
e(u-\theta_j)e(u+\theta_j) \frac{Q^{(1)}(u+3)}{Q^{(1)}(u+2)}
h_2(u+3)\tilde{h}_2(u+3),\no\\[4pt]
&&Q^{(m)}(u)=\prod_{k=1}^{L_m}(u-\lambda_k^{(m)}+\frac{m}{2})(u+\lambda_k^{(m)}+\frac{m}{2}),\quad m=1,2, \no \\[4pt]
&&f_1(u)=\frac{1}{ 2^{2}} \frac{u
(u+1)(u+3)}{(u+\frac{3}{{2}}) }\prod_{j=1}^N
b(u-\theta_j)b(u+\theta_j)(u-\theta_j+1)(u+\theta_j+1)\no\\[4pt]
&&\hspace{20mm}\times \frac{Q^{(2)}(u-\frac{1}{{2}})Q^{(2)}(u-\frac{3}{{2}})}{Q^{(1)}(u)}{h_1(u)\tilde{h}_1(u)} x,\no\\[4pt]
&&f_2(u)=\frac{1}{ 2^{2}} \frac{u (u+2)(u+3)}{(u+\frac{3}{{2}})
}\prod_{j=1}^N
b(u-\theta_j)b(u+\theta_j)(u-\theta_j+2)(u+\theta_j+2)\no\\[4pt]
&&\hspace{20mm}\times \frac{Q^{(2)}(u+\frac{3}{{2}})Q^{(2)}(u+\frac{5}{{2}})}{Q^{(1)}(u+2)}{h_2(u+3)\tilde{h}_2(u+3)} \,x,\label{Function-2}
\eea where $x$ is a constant  related to the boundary parameters (see below (\ref{x-parameter})) and $\{h_i(u),\,\tilde{h}(u)|i=1,2\}$ are
some functions given by
\bea
&&h_1(u)=2(\sqrt{1+c_1c_2}u+\zeta), \qquad  h_2(u)=2(\sqrt{1+c_1c_2}u-\zeta), \no\\[4pt]
&&\tilde{h}_1(u)=-2(\sqrt{1 +\tilde{c}_1\tilde{c}_2}u-\tilde{\zeta}), \quad \tilde{h}_2(u)=-2(\sqrt{1 +\tilde{c}_1\tilde{c}_2}u+\tilde{\zeta}).
\eea
The eigenvalues $\Lambda(u)$ and $\bar \Lambda(u)$ can be expressed as\footnote{It is well-known that the $T-Q$ relation  for eigenvalues is not unique and there exist many
representations \cite{Cao2} which all give rise to the same set of eigenvalues. Hence we present the $T-Q$ relation (\ref{eop-1})-(\ref{eop-2321}) which leads to
the whole set of eigenvalues of the model.}
\bea
&&\Lambda (u)=Z_1(u)+
Z_2(u)+Z_3(u)+Z_4(u)+f_1(u)+f_2(u),\label{eop-1} \\[4pt]
&&\bar \Lambda(u)=\prod_{i=1}^N[(u-\theta_i+\frac32)(u+\theta_i+\frac32)\tilde{\rho}_0(u-\theta_i+\frac12)\tilde{\rho}_0(u+\theta_i+\frac12)]^{-1} \no\\[4pt]
&&\qquad \quad\times \frac{1}{2^4} \rho_1(2u+3)
(u-\frac12)^{-1} (u+\frac32)^{-2} (u+\frac72)^{-1}\no\\[4pt]
&&\qquad \quad\times
\{Z_1(u+\frac12)[Z_2(u-\frac12)+Z_3(u-\frac12)+Z_4(u-\frac12)+f_2(u-\frac12) ]\no\\[4pt]
&&\qquad \quad
+[Z_2(u+\frac12)+Z_3(u+\frac12)+f_1(u+\frac12)]Z_4(u-\frac12)+Z_2(u+\frac12)f_2(u-\frac12)\no\\&&\qquad\quad
+f_1(u+\frac12)Z_3(u-\frac12)+f_1(u+\frac12)f_2(u-\frac12)\}, \label{eop-2321} \eea
where the non-negative integers $L_1$ and $L_2$ satisfy
\bea
L_1=2 L_2+N+1.\no\eea
Because the eigenvalues
$\Lambda(u)$ and $\bar \Lambda(u)$ are both  polynomials of $u$, the regularity of these functions gives the constraints of Bethe roots
\bea &&\frac{(\lambda_k^{(1)}+\frac{1}{2})}{\lambda_k^{(1)}}
\frac{1}
{\prod_{j=1}^N(\lambda_k^{(1)}-\theta_j-\frac{1}{2})(\lambda_k^{(1)}+\theta_j-\frac{1}{2})}
\frac{Q^{(1)}(\lambda_k^{(1)}-\frac{3}{2})}{Q^{(2)}(\lambda_k^{(1)}-2)}\no\\[4pt]
&&\qquad\qquad +\frac{(\lambda_k^{(1)}-\frac{1}{2})}{\lambda_k^{(1)}}
\frac{1}{\prod_{j=1}^N(\lambda_k^{(1)}-\theta_j+\frac{1}{2})
(\lambda_k^{(1)}+\theta_j+\frac{1}{2})}
\frac{Q^{(1)}(\lambda_k^{(1)}+\frac{1}{2})}{Q^{(2)}(\lambda_k^{(1)})}\no\\[4pt]
&&\qquad\qquad+x(\lambda_k^{(1)}+\frac{1}{2})(\lambda_k^{(1)}-\frac{1}{2}){Q^{(2)}(\lambda_k^{(1)}
-1)}=0, \quad k=1,2,\cdots,L_1, \label{opba-1} \\[6pt]
&&\frac{1}{(\lambda_l^{(2)}-1)}
\frac{{Q^{(2)}(\lambda_l^{(2)}-3)}}{Q^{(1)}(\lambda_l^{(2)}-\frac{3}{2})}
h_1(\lambda_l^{(2)}-\frac{3}{2})\tilde{h}_1(\lambda_l^{(2)}-\frac{3}{2})\no\\[4pt]
&&\qquad\qquad+\frac{1}{(\lambda_l^{(2)}\hspace{-0.12truecm}+\hspace{-0.12truecm}1)}
\frac{{Q^{(2)}(\lambda_l^{(2)}\hspace{-0.12truecm}+\hspace{-0.12truecm}1)}}{Q^{(1)}(\lambda_l^{(2)}\hspace{-0.12truecm}+\hspace{-0.12truecm}\frac{1}{2})}
h_2(\lambda_l^{(2)}\hspace{-0.12truecm}+\hspace{-0.12truecm}\frac{3}{2})\tilde{h}_2(\lambda_l^{(2)}\hspace{-0.12truecm}+\hspace{-0.12truecm}\frac{3}{2})
=0,\,l=1,2,\cdots,L_2. \label{opba-2}\eea
We note that the Bethe
ansatz equations obtained from the regularity of $\Lambda(u)$ are  the same as those obtained from the regularity of $\bar \Lambda(u)$.
The function $Q^{(m)}(u)$ has two zero points, namely, $\lambda_k^{(m)}-\frac m2$ and
$-\lambda_k^{(m)}-\frac m2$. We have checked that the Bethe ansatz equations obtained from
these two points are the same. $x$ is fixed by the asymptotic behaviors of $\Lambda(u)$ and
$\bar \Lambda(u)$ as
\bea
x=\frac{2+c_1\tilde{c}_2+c_2\tilde{c}_1}{\sqrt{(1+c_1c_2)(1 +\tilde{c}_1\tilde{c}_2)}}-2.\label{x-parameter}\eea
If $c_1=c_2=\tilde{c}_1=\tilde{c}_2=0$, the boundary reflection matrices reduce to diagonal ones. In this case, $x=0$ and our solution of $\Lambda(u)$ (after taking the homogeneous limit $\{\theta_j\to0|j=1,2,\cdots,N\}$) naturally reduces to that obtained via algebraic Bethe method \cite{c2aba}. Finally, the eigenvalue $E$ of Hamiltonian (\ref{hh}) can be expressed in terms of the Bethe roots as
\begin{eqnarray}
E= \frac{1}{2}\frac{\partial \ln \Lambda(u)}{\partial
u}|_{u=0,\{\theta_j\}=0}.\no
\end{eqnarray}

\section{Discussion}

In this paper, we generalize the nested ODBA method to the integrable models related to the $sp(4)$ Lie algebra.
By using the fusion technique, we obtain the closed operator product identities of the fused transfer matrices. Based on them and the asymptotic
behaviors as well as the values of transfer matrices at some special points of $u$, we obtain the exact solution of the system with the periodic and off-diagonal open boundary conditions.
The method developed in this paper can be generalized to the high rank $C_n$ (i.e., the $sp(2n)$) case directly\footnote{Although the method in this paper can be generalized
to the high rank $C_n$ straightforward, it still takes some tedious calculations.}.

\section*{Acknowledgments}

The financial supports from the National Program
for Basic Research of MOST (Grant Nos. 2016YFA0300600 and
2016YFA0302104), the National Natural Science Foundation of China
(Grant Nos. 11434013, 11425522, 11547045, 11774397, 11775178 and 11775177), the Major Basic Research Program of Natural Science of Shaanxi Province
(Grant Nos. 2017KCT-12, 2017ZDJC-32), Australian Research Council (Grant No. DP 190101529) and the Strategic Priority Research Program of the Chinese
Academy of Sciences, and the Double First-Class University Construction Project of Northwest University are gratefully acknowledged.
GL Li would like to thank Shaanxi Province Key
Laboratory of Quantum Information and Quantum Optoelectronic
Devices, Xian Jiaotong University for its support. The authors also would like to thank F. Wen, P. Sun and Y. Qiao for their valuable discussions.



\begin{thebibliography}{99}

\bibitem{1} R. J. Baxter, {\it Exactly Solved Models in Statistical Mechanics},
Academic Press, 1982.
\bibitem{2} C. N. Yang, {\it Phys. Rev. Lett.} {\bf 19} (1967), 1312.

\bibitem{Alc87} F. C. Alcaraz, M. N. Barber, M. T. Batchelor,
R. J. Baxter and G. R. W. Quispel, {\it J. Phys. A} {\bf 20} (1987),
6397.
\bibitem{Skl88} E. K. Sklyanin, {\it J. Phys. A} {\bf 21} (1988), 2375.

\bibitem{Kor93} V.\,E. Korepin, N.\,M. Bogoliubov and A.\,G. Izergin,
{\it Quantum Inverse Scattering Method and Correlation Function\/},
Cambridge University Press, 1993.
\bibitem{Cao03} J. Cao, H. -Q. Lin, K. -J. Shi and Y. Wang,
Nucl. Phys. B {\bf 663}, 487 (2003).
\bibitem{Nepo02}R. I. Nepomechie, Nucl. Phys. B {\bf 622}, 615 (2002).
\bibitem{Bas1} P. Baseilhac, {\it Nucl. Phys. B} {\bf 754} (2006), 309.
\bibitem{Bas2} P. Baseilhac and K. Koizumi, {\it J. Stat. Mech.} (2007),
P09006.
\bibitem{Fra08} H. Frahm, A. Seel and T. Wirth, {\it Nucl. Phys. B} {\bf 802} (2008), 351.
\bibitem{Bas3} P. Baseilhac and S. Belliard, {\it Lett. Math. Phys.} {\bf 93} (2010), 213.
\bibitem{Bas31} P. Baseilhac and S. Belliard, {\it Nucl. Phys. B} {\bf 873} (2013), 550.
\bibitem{Nep13} R. I. Nepomechie, {\it J. Phys. A} {\bf 46} (2013), 442002.
\bibitem{Bel13} S. Belliard and N. Cramp{\'e}, {\it SIGMA} {\bf 9} (2013), 072.
\bibitem{Bel15} S. Belliard, {\it Nucl. Phys. B} {\bf 892} (2015), 1.
\bibitem{Bel15-1} S. Belliard  and R.\,A. Pimenta, {\it Nucl. Phys. B} {\bf 894} (2015), 527.
\bibitem{Ava15} J. Avan, S. Belliard, N. Grosjean and R.\,A. Pimenta, {\it Nucl. Phys. B} {\bf 899} (2015), 229.



\bibitem{Cao1} J. Cao, W.-L. Yang, K. Shi and Y. Wang, {\it Phys. Rev. Lett.} {\bf 111} (2013),
137201.
\bibitem{Cao2} Y. Wang, W. -L. Yang, J. Cao and K. Shi, {\it Off-Diagonal Bethe
Ansatz for Exactly Solvable Models}, Springer Press, 2015.
\bibitem{Zha13} X. Zhang, J. Cao, W.-L. Yang, K. Shi and Y. Wang, {\it J. Stat. Mech.} (2014), P04031.
\bibitem{Hao14} K. Hao, J. Cao, G. -L. Li, W. -L. Yang, K. Shi and Y. Wang,
{\it JHEP} {\bf 06} (2014), 128.




\bibitem{Cao14JHEP143}J. Cao, W.\,-L. Yang, K. Shi and Y. Wang,  {\it JHEP} {\bf 04} (2014), 143.

\bibitem{NYRes} N. Yu. Reshetikhin,  {\it Sov. Phys. JETP.} {\bf 57} (1983),
691.
\bibitem{Bn} M. J. Martins and P. B. Ramos, {\it Nucl. Phys. B} {\bf 500} (1997), 579.


\bibitem{Kar79-1} M. Karowski, {\it Nucl. Phys. B} {\bf 153} (1979), 244.
\bibitem{Kar79-10} P. P. Kulish, N. Yu. Reshetikhin and E. K. Sklyanin, {\it Lett. Math. Phys.} {\bf
5} (1981), 393.
\bibitem{Kar79-2} P. P. Kulish and E. K. Sklyanin, {\it Lecture Notes
in Physics} {\bf 151} (1982), 61.
\bibitem{Kar79-3} A. N. Kirillov and N.Yu.
Reshetikhin, {\it J. Sov. Math.} {\bf 35} (1986), 2627; {\it J. Phys. A} {\bf 20} (1987),
1565.
\bibitem{Mez92} L. Mezincescu and  R. I. Nepomechie, {\it Nucl. Phys. B} {\bf 372} (1992),
597.
\bibitem{Zho96} Y. -K. Zhou, {\it Nucl. Phys. B} {\bf 458} (1996), 504.




\bibitem{8Mel054} H.\,J. de Vega and A. Gonz$\acute{a}$lez-Ruiz, {\it Nucl. Phys. B} {\bf 417} (1994), 553.
\bibitem{8Mel055} H.\,J. de Vega and A. Gonz$\acute{a}$lez-Ruiz, {\it Mod. Phys. Lett. A} {\bf 09} (1994), 2207.
\bibitem{8Mel056} G. -L. Li, R. H. Yue and B. Y. Hou, {\it Nucl. Phys. B} {\bf 586} (2000),
711.


\bibitem{c2aba} G. -L. Li, K. J. Shi and R. H. Yue, {\it Commun. Theor. Phys. } {\bf 44} (2005),
89.






















\end{thebibliography}
\end{document}